\documentclass[11pt,a4paper]{article}
\pdfoutput=1
\usepackage{jheppub}
\usepackage{slashed}
\usepackage{cancel}
\usepackage{xcolor}

\makeatletter
\def\@fpheader{\relax}
\makeatother

\usepackage{ulem}

\usepackage[czech,english]{babel}
\usepackage{graphicx}
\usepackage{yfonts}
\usepackage{amsmath,amsfonts,amssymb}
\usepackage{url}
\usepackage{mathabx}
\usepackage{ulem}
\usepackage{mathtools}

\DeclareMathOperator{\MyProd}{\scalebox{1.4}{$\mathrm{I\kern-0.2ex I}$}}

\preprint{LCTP-21-14}

\title{Logarithmic Corrections to the Entropy of Rotating Black Holes and Black Strings  in AdS$_5$}

\author[a]{Marina David}

\emailAdd{mmdavid@umich.edu}

\author[b, c, d, e]{Alfredo Gonz\'alez Lezcano}

\emailAdd{agonzale@ictp.it}

\author[f, g]{Jun Nian}

\emailAdd{Jun.Nian@mi.infn.it}

\author[a, b]{Leopoldo A. Pando Zayas}

\emailAdd{lpandoz@umich.edu}

\affiliation[a]{Leinweber Center for Theoretical Physics, University of Michigan, Ann Arbor, MI 48109, USA}
\affiliation[b]{The Abdus Salam International Centre for Theoretical Physics, 34014 Trieste, Italy}
\affiliation[c]{S.I.S.S.A., International School for Advanced Studies, 34136 Trieste, Italy}
\affiliation[d]{INFN, Sezione di Trieste, 34127 Trieste, Italy}
\affiliation[e]{ Asia Pacific Center for Theoretical Physics, Postech, Pohang 37673, South Korea}
\affiliation[f]{INFN, Sezione di Milano, 20133 Milano, Italy}
\affiliation[g]{International Centre for Theoretical Physics Asia-Pacific, University of Chinese Academy of\\ Sciences, 100190 Beijing, China}

\abstract{We investigate logarithmic corrections to the entropy of supersymmetric, rotating, asymptotically AdS$_5$ black holes and black strings. Within the framework of the AdS/CFT correspondence, the entropy of these black objects is determined,  on the field theory side, by  the superconformal index and  the  refined topologically twisted index of ${\cal N}=4$ supersymmetric Yang-Mills theory, respectively. We read off the logarithmic correction from those field-theoretic partition functions.   On the gravity side, we take the near-horizon limit and apply the Kerr/CFT correspondence  whose associated charged Cardy formula describes the degeneracy of states at subleading order  and determines the logarithmic correction to the entropy. We find  perfect agreement between these two approaches.  Our results provide a window into  precision microstate counting and demonstrate the efficacy of  low-energy, symmetry-based approaches such as the Kerr/CFT correspondence for asymptotically AdS black objects under certain conditions.}


\arxivnumber{}

\newcommand{\bea}{\begin{eqnarray}}
\newcommand{\eea}{\end{eqnarray}}

\newcommand{\be}{\begin{equation}}
\newcommand{\ee}{\end{equation}}

\begin{document}

\maketitle

\section{Introduction}
One of the fundamental results in the study of black holes during the 1970's was to establish that they can be attributed an entropy \cite{Bekenstein:1973ur, Hawking:1974sw}, which further fits into a set of  rules similar to the thermodynamic laws \cite{Bardeen:1973gs, Bekenstein:1974ax, Hawking:1976de}. At the leading order in Newton's constant, the entropy was found to be universally given by a quarter of the area of the horizon. When further quantum effects are  taken into consideration, logarithmic corrections to the entropy arise.

Entropy is an important bridge between the macroscopic and the microscopic  worlds, and we expect any candidate quantum theory of gravity to provide an explanation for the macroscopic Bekenstein-Hawking entropy in terms of microscopic counting of degrees of freedom. As a leading candidate for a theory of quantum gravity, string theory has already provided, in the work  of  Strominger and Vafa, a microstate counting of the entropy  for certain asymptotically flat black holes \cite{Strominger:1996sh}. It has also provided, through work by Sen and collaborators, an explanation for the logarithmic corrections to the entropy \cite{Sen:2014aja}.

The AdS/CFT correspondence, stating a mathematical equivalence between certain theories containing gravity in AdS and field theories on the boundary of AdS \cite{Maldacena:1997re}, potentially translates intricate questions of black hole dynamics into questions in unitary field theories.  Using the AdS/CFT correspondence,  Benini, Hristov and Zaffaroni have given a microscopic foundation to the entropy of certain magnetically charged AdS$_4$ black holes in terms of a field theory counting of microstates  \cite{Benini:2015eyy}. More recently, microscopic foundations in terms of field-theoretic partition functions have been provided for rotating, electrically charged, asymptotically AdS black holes \cite{Cabo-Bizet:2018ehj, Choi:2018hmj, Benini:2018ywd,Choi:2019miv,  Kantor:2019lfo,Nahmgoong:2019hko,Choi:2019zpz,  Nian:2019pxj}. For some of these black holes it has been shown that the field-theoretic, microscopic correction to the entropy matches precisely the macroscopic logarithmic corrections arising from one-loop supergravity  \cite{Liu:2017vbl, Gang:2019uay, Benini:2019dyp, PandoZayas:2020iqr}.

The recent developments use the full UV complete description of gravity in AdS$_{d+1}$, the supersymmetric field theories in $d=3,4,5,6$. Guided by the principle of renormalization group (RG) flow, one naturally expects that a universal property such as the black hole entropy can be recovered from an effective theory without full knowledge of the UV complete descripion, as verified  in previous cases \cite{Strominger:1997eq}. For AdS black holes, applying the Kerr/CFT correspondence \cite{Guica:2008mu,Lu:2008jk, Chow:2008dp}, a unified microscopic description of the entropy of asymptotically AdS$_{4,5,6,7}$ black holes was given in  \cite{David:2020ems}  for extremal and extended in  \cite{Nian:2020qsk, David:2020jhp} to near-extremal black holes.

The goal of this paper is to explore logarithmic corrections to the entropy of AdS$_5$ black holes and black strings from both the UV complete boundary $\mathcal{N}=4$ SYM theory and the near-horizon 2-dimensional CFT.  From the UV complete field theoretical point of view, the observables are the topologically twisted index and the superconformal index, both of which are initially given in the grand-canonical ensemble. Only after a Legendre transform to the microcanonical ensemble does the index probe the degeneracies of states giving rise to black hole entropy \cite{Cabo-Bizet:2018ehj, Murthy:2020rbd}. We should also point out that the Kerr/CFT correspondence, being entirely based on the near horizon geometry, is not suitable to probe the entropy of the black hole in the grand-canonical ensemble since the values of the chemical potentials are to be fixed at the boundary of the full geometry. Hence, we probe the black hole entropy at fixed charge and angular momentum \cite{Pathak:2016vfc} in the microcanonical ensemble, and compare the results from the two approaches in this ensemble\footnote{Further discussion about the entropy of black hole in different ensembles can be found in \cite{Sen:2014aja, Sen:2012dw}}.

We  review the logarithmic corrections as arising from the analysis of the large-$N$ limit of the corresponding partition functions responsible for the UV complete counting of the entropy for the black holes and black strings in AdS$_5$. Namely, we consider the logarithmic corrections to the superconformal index and the topologically twisted index of ${\cal N}=4$ SYM. We  also exploit the fact that in the Kerr/CFT approach the degeneracy of states is given by the charged Cardy formula, which can and has been evaluated to the subleading order that contains logarithmic corrections  \cite{Carlip:2000nv,Sen:2012dw,Pathak:2016vfc}. Logarithmic corrections to the entropy can be computed macroscopically from  one-loop contributions of the massless sector in the corresponding supergravity theory  \cite{Banerjee:2010qc, Banerjee:2011jp, Sen:2011ba}, and therefore provide a litmus test to any UV complete description.  We find that both approaches yield the same logarithmic correction in the microcanonical ensemble.

This paper is organized as follows. In Sec.~\ref{sec:AdS5BH} we review the AdS$_5$ black hole entropy from the superconformal index of the boundary $\mathcal{N}=4$ SYM and from the near-horizon Kerr/CFT, and then compute the logarithmic correction to the entropy using both approaches. In Sec.~\ref{sec:AdS5BS} we apply similar techniques and use two approaches to compute the logarithmic correction to the entropy of rotating AdS$_5$ black strings. A brief summary and some proposals for the future research are presented in Sec.~\ref{sec:discussion}. Some relevant facts about the special functions are collected in Appendix~\ref{app:SpecialFunctions}.

\section{AdS$_5$ Black Holes}\label{sec:AdS5BH}

\subsection{The Superconformal Index and Black Hole Entropy} \label{sec:thesci}

An efficient way to count $\frac{1}{16}$-BPS states in ${\cal N}=4$ SYM is to consider the theory on $S^1\times S^3$ and evaluate the superconformal index (SCI) \cite{Romelsberger:2005eg, Kinney:2005ej}:  
\begin{align}
    \mathcal{I}(\tau; \Delta) & = \text{Tr}\left[\left(-1\right)^{F}e^{- \beta \{\mathcal{Q}, \mathcal{Q}^{\dagger}\}}v_a^{Q_a}p^{J_1 + \frac{r}{2}} q^{J_2 + \frac{r}{2}}\right], \label{Eq:TheSCI}
\end{align}
where $\beta$ is the circumference of $S^1$, and $F$ is the fermionic operator, while $Q_{a=1,2,3}$ are flavor charges with associated fugacities $v_a =e^{2 \pi i \Delta_a}$. With $r$ we denote the R-charge. The fugacities $p=e^{2 \pi i \tau}$ and  $q=e^{2 \pi i \sigma}$ are associated to the angular momenta $J_{1,2}$ of $S^3$, and the combinations $J_{1,2}+\frac{r}{2}$ commute with the supercharge $\mathcal{Q}$. In what follows we set $\tau =\sigma$ for simplicity. Note that the counting of states that the SCI offers should be seen as performed in the grand-canonical ensemble, since we are keeping fixed the values of chemical potentials while summing over all possible charges.

According to the AdS/CFT correspondence, $SU(N)$ $\mathcal{N}=4$ SYM is dual to type IIB supergravity on AdS$_5 \times S^5$, in which one can find supersymmetric black hole solutions that are asymptotically AdS, rotating and electrically charged. Remarkably, in recent years plenty of evidences have been gathered indicating that $\mathcal{I}(\tau; \Delta)$ captures the entropy of such black holes \cite{Cabo-Bizet:2018ehj, Choi:2018hmj, Benini:2018ywd} (see \cite{Lezcano:2019pae,Lanir:2019abx,Ardehali:2014zba,Amariti:2019mgp,Ardehali:2015bla,DiPietro:2016ond, Benini:2020gjh} for further developments and \cite{GonzalezLezcano:2020yeb, GonzalezLezcano:2021ycc} for a more complete list of references). 

The SCI can be written as a contour integral over the holonomies of the gauge group \cite{Romelsberger:2007ec,Dolan:2008qi}:
\begin{align}
\begin{split}\label{Eq:indexN4}
\mathcal{I} \left(\tau; \Delta\right) & = \kappa_N\int_{0}^1\prod_{\mu =1}^{N-1}d u_{\mu}
\mathcal{Z}\left(u; \Delta, \tau \right)\, ,\\
\mathcal{Z}\left(u; \Delta, \tau \right) & = \frac{\prod_{a=1}^{3}\prod_{i \neq j}\widetilde{\Gamma} \left(u_{ij}+\Delta_a;\tau\right)}{\prod_{i \neq j}\widetilde{\Gamma} \left(u_{ij}; \tau\right) }\, ,\\
\kappa_{N} & = \frac{\left(p ; p \right)_{\infty}^{N-1}\left(q ; q \right)_{\infty}^{N-1}}{N!} \prod_{a=1}^{3}\left(\widetilde{\Gamma}(\Delta_a; \tau)\right)^{N-1}\, ,
\end{split}
\end{align}
where $(\cdot\, ; \cdot)_{\infty}$ is the Pochhammer symbol, and $\widetilde{\Gamma}(u; \tau)$ is the elliptic Gamma function defined both in Appendix~\ref{app:SpecialFunctions}. 
There are two main approaches to evaluate the $N$-dimensional integral over the holonomies of the gauge group representing the SCI.  The first approach relies on a direct application of the residue theorem, and yields what is known in the literature as the Bethe-Ansatz method. The second approach implements a saddle-point evaluation of the integral.

\subsubsection{The Bethe-Ansatz Approach}
The location of the poles of \eqref{Eq:indexN4} is given by the solutions to the set of equations:
\begin{equation}
\begin{split}
Q_k(\hat{u}; \Delta, \tau) =1, \hspace{2mm} \forall \hspace{1.5mm} k =1, \cdots, N\, ,\label{BAEs} 
\end{split}
\end{equation}
where 
\begin{equation}
Q_k(u; \Delta, \tau) = e^{2 \pi i \lambda} \prod_{l=1 ( \neq k)}^{N}\prod_{a =1}^3 \frac{ \theta_1(- u_{kl}+\Delta_a; \tau )}{\theta_1(u_{kl}+\Delta_a; \tau)} \label{Eq:BAOperator}
\end{equation}
are the Bethe-Ansatz operators and the values $\hat{u}$ satisfying \eqref{BAEs} are called Bethe-Ansatz solutions. We then define $\text{BA}= \left\{\hat{u}\hspace{2mm}| \hspace{2mm}\eqref{BAEs} \hspace{2mm} \text{is satisfied}\right\}$. With  $\lambda$ we have denoted a Lagrange multiplier implementing the $SU(N)$ constraint  on the holonomies $\sum_{i=1}^N u_i \in \mathbb{Z}$,  and $\theta_1(u;\tau)$ is the elliptic theta function defined in Appendix~\ref{app:SpecialFunctions}.
Upon direct application of the residue theorem, $\mathcal{I}(\tau, \Delta)$ can be rewritten in terms of a discrete sum as: 
\begin{equation}
\begin{split}
	\mathcal I(\tau; \Delta)&=\kappa_N\sum_{\hat u\in\mathrm{BA}}\mathcal Z(\hat u;\Delta,\tau)H(\hat u;\Delta,\tau)^{-1}\, ,\label{eq:index:BA}\\
H\left(\hat{u}; \Delta, \tau\right)& = \det \left[\frac{1}{2 \pi i} \frac{\partial \left(Q_1, \cdots, Q_N\right)}{\partial \left(u_1, \cdots , u_{N-1}, \lambda\right)}\right]\, .
	\end{split}
\end{equation}

Let us emphasize that \eqref{eq:index:BA} is not the full story, since the application of the residue theorem required the poles to be isolated, and there is enough evidence by now  \cite{Lezcano:2021qbj,Benini:2021ano} that this is not the case generically. We shall focus only on the contributions coming from isolated poles (see \cite{Hong:2021bzg} for more detailed discussions on this point). A set of solutions  to the equations \eqref{BAEs} was found in \cite{Hong:2018viz} and it is given by: 
  \begin{align}
  \begin{split}
      u_i & = u_{\hat{j},\hat{k}} = \Bar{u} + \frac{\hat{j}}{m} + \frac{\hat{k}}{n}\left(\tau +\frac{r}{m}\right)\, ,  \label{eq:BAsolTT}\\
      \hat{j}& = 0, \cdots, m-1\, , \qquad \hat{k} = 0, \cdots, n-1\, ,\\
r & =0,\cdots, n-1\, ,
      \end{split}
  \end{align}
  where $N= m n$, hence, each set $\{u_i\}$ in \eqref{eq:BAsolTT} can be labeled by the numbers $\{m,n,r\}$. These solutions to the Bethe-Ansatz equations for the SCI were, in fact,  inspired by the set of solutions found in \cite{Hong:2018viz} for the Bethe-Ansatz equations associated to the topologically twisted index. We will discuss that case in Sec.~\ref{sec:blackstring}. However, in the large-$N$ limit, it was possible to argue that the configuration corresponding to $\{1,N,0\}$ contributed dominantly to the SCI. We shall refer to the $\{1,N,0\}$ solution as the ``basic'' solution, namely
\begin{equation}
	\hat u_{\text{basic}}=\left\{u_i=\bar u+\frac{i}{N}\tau\,\Big|\,i=1,2,\cdots,N-1\right\}\bigcup\left\{u_N=\bar u\right\}\, ,\label{BAE:sol:basic}
\end{equation}
where $\bar u$ is determined as
\begin{equation}
	N\bar u+\frac{N(N-1)}{2N}\tau\in\mathbb Z\, .\label{ubar:general}
\end{equation}
 The parameter $\Bar{u}$ enforces the $SU(N)$ constraint $\sum_{i=1}^N u_i \in \mathbb{Z}$ on the holonomies, which, together with the periodicity properties of $\mathcal{I}(\tau; \Delta)$ allows us to obtain:
 \begin{align}
     \Bar{u} & = \frac{k}{N} - \frac{N-1}{2N} \tau\, , \qquad k=1,\cdots, N-1\, , \label{eq:ubar}
 \end{align}
each of which contributes identically to the $\mathcal{I}(\tau; \Delta)$, we therefore have that, in the appropriate regime of chemical potentials:

\begin{equation}
\begin{split}
	\log \mathcal I(\tau; \Delta)\big|_\text{Basic BA}& =-\frac{i \pi (N^2 -1)}{ \tau^2}\Delta_1 \Delta_2 \Delta_3 +\log N\,+\mathcal O(N^0). \label{eq:SCIBA}
\end{split}
\end{equation}
We see that the $\log N$ in \eqref{eq:SCIBA} has a purely combinatorial origin, whose precise form is quite insensitive to details about the theory in which $\mathcal{I}(\tau; \Delta)$ is being evaluated.

\subsubsection{The Saddle Point Approach and the Cardy-Like Expansion}
Let us further reinforce the idea that the logarithmic correction to the SCI has a combinatorial origin. To do so we briefly reproduce here the saddle-point evaluation of \eqref{Eq:indexN4} implemented in \cite{GonzalezLezcano:2020yeb}. 

\subsubsection*{The strict Cardy-like limit}
By the strict Cardy-like limit we mean that we keep only the most divergent term in a $\tau \rightarrow 0$ expansion\footnote{More refined limits were considered in \cite{ArabiArdehali:2021nsx}. The authors discussed the limit where $q=e^{2 \pi i \tau}$ approaches roots of unit.}.  The study of the strict Cardy-like limit was the subject of several works \cite{Choi:2018hmj, Cabo-Bizet:2019osg, Honda:2019cio, ArabiArdehali:2019tdm, Kim:2019yrz, Amariti:2019mgp}, and the main idea is to rewrite \eqref{Eq:indexN4} in the following way:
\begin{align}
\begin{split}
\mathcal{I}(\tau; \Delta) & = \kappa_N\int\prod_{\mu =1}^{N-1}d u_{\mu}
\exp \left( \frac{1}{\tau^2}S_{\text{eff}}(u; \Delta, \tau)\right)\, ,
\end{split}
\end{align}
where $S_{\text{eff}}(u; \Delta, \tau)$ is appropriately defined such that $\mathcal{Z}(u;\Delta, \tau)$ in \eqref{Eq:indexN4} is recovered. The $\frac{1}{\tau^2}$ factor can be used as a large control parameter to apply the saddle-point method in the strict Cardy-like limit. We are exploiting the fact that we already know the leading contribution in such limit is precisely of the order $\mathcal{O}\left(\frac{1}{\tau^2}\right)$.  The saddle-point equations have the form:

\begin{equation}
	\frac{\partial}{\partial u_\mu}S_\text{eff}( u;\Delta,\tau)\bigg|_{u=\text{saddle}} = 0\, , \quad\quad (\mu=1,\cdots,N-1)\, .\label{eq:saddlep}
\end{equation}

The set with all identical holonomies, namely $u_i=u_j$ for all $i,j\in\{1,\cdots,N\}$ \cite{Choi:2018hmj,Honda:2019cio} is one of the most well-known solutions to \eqref{eq:saddlep}. The effective action at this saddle point successfully counted the dual AdS$_5$ black hole microstates \cite{Choi:2018hmj}.

There are $N$ distinct sets of identical holonomies satisfying the $SU(N)$ constraint $\sum_{i=1}^Nu_i\in\mathbb Z$, namely
\begin{equation}
	 u^{(m)}=\left\{u_j^{(m)}=\frac{m}{N}\,\Big|\,j=1,\cdots,N\right\}\, ,\qquad(m=0,1,\cdots,N-1)\, .
\label{eq:saddle:sol}
\end{equation}
Within the appropriate range of chemical potentials, the saddle points \eqref{eq:saddle:sol} yield the following effective action:
\begin{align}
\frac{1}{\tau^2}\sum_{m=0}^{N-1}S_{\text{eff}}(u^{(m)}; \Delta, \tau) & = \exp \left(-\frac{i \pi (N^2 -1)}{ \tau^2}\Delta_1 \Delta_2 \Delta_3 +\log N\,+\mathcal O(-1/|\tau|)\right)\, . \label{eq:sadeval}
\end{align}
From \eqref{eq:saddle:sol} and \eqref{eq:sadeval} we see that the logarithmic correction has its origin in the multiplicity of the saddle points. This result remains true even for more generic $\mathcal{N}=1$ toric quiver gauge theories, as emphasized in \cite{GonzalezLezcano:2020yeb}, which renders the $\log N$ correction a quite robust one. Note that we have not made use of the large-$N$ limit here, therefore, provided that we remain at small values of $\tau$, \eqref{eq:sadeval} holds for finite $N$ (Evidence in favor of this has been given in \cite{Lezcano:2021qbj}).

\subsubsection*{The Cardy-like expansion}

With the Bethe-Ansatz approach we have learned that even for generic values of $\tau$, in the large-$N$ limit, the $\log N$ is the same and arises from degeneracies of Bethe-Ansatz solutions.
At this point we have shown that also for finite $N$, in the strict Cardy-like limit, the $\log N$ has a combinatorial origin. 

 We now proceed to include subleading corrections in inverse powers of $\tau$ and show that, indeed, the $\log N$ remains unchanged. This is an important step, since it helps us build an intuition that we later import to a different situation, namely the refined topologically twisted index, where we have only access to the strict Cardy-like limit and argue about the possibility of the combinatorial nature of $\log N$ to remain true as we depart from this limit.
We then focus on the effective action  evaluated near the leading saddle-point solution (\ref{eq:saddle:sol}). Following \cite{GonzalezLezcano:2020yeb}, we make the Ansatz for saddle-point solutions in the finite Cardy-like expansion,
\begin{equation}
	 u^{(m)}=\left\{u_j^{(m)}=\frac{m}{N}+v_j\tau\,\Big|~v_j\sim\mathcal O(|\tau|^0),~\sum_{j=1}^Nv_j=0\right\}\, ,\quad(m=0,1,\cdots,N-1)\, ,\label{eq:saddle:ansatz:finite}
\end{equation}
and evaluate the effective action around this Ansatz. For a suitable choice of chemical potentials, the following expression was obtained:

\begin{equation}
\begin{split}
	\log \mathcal I(\tau; \Delta)& =-\frac{i \pi (N^2 -1)}{ \tau^2}\Delta_1 \Delta_2 \Delta_3 +\log N\,+\mathcal O(e^{-1/|\tau|})\, . \label{eq:SCI}
\end{split}
\end{equation}
The exponentially suppressed correction comes from the asymptotic expansion of the building blocks of the effective action, namely elliptic Gamma functions  and Pochhammer symbols (see Appendix \ref{app:assypmbehavior}). An important aspect about \eqref{eq:SCI} is that it includes all power-like corrections in $\tau$, and rather remarkably, it is a series that truncates at the leading order. A prominent role in the technical evaluation of \eqref{eq:SCI} was the cancellation of the $\mathcal{O}(\tau^0)$  contribution which was given in terms of the effective action of a matrix model of $SU(N)$ level $k=N$ Chern-Simons theory on $S^3$ (see \cite{GonzalezLezcano:2020yeb} for a more detailed discussion).

What seems like a rather technical step when analyzed from the strictly mathematical perspective of looking at the asymptotic behavior of $S_{\text{eff}}(u; \Delta,\tau)$, becomes very natural when viewed from an effective field theory perspective. Such analysis was carried out  in \cite{Cassani:2021fyv}, where the Cardy-like (small $\tau$) expansion  was shown to geometrically correspond to shrinking the $S^1$ circle, thus leading to an effective field theory on $S^3$ organized in inverse powers of the circumference of $S^1$. 

In particular, a careful treatment of the Kaluza-Klein reduction on $S^1$ yields a result compatible with \eqref{eq:SCI}, where the $\log N$ is associated to degeneracies of vacua. This effective field theory approach clarifies the organization of the index in inverse powers of $|\tau|$ and further confirms the logarithmic term as certain degeneracy of vacua \cite{Cassani:2021fyv,ArabiArdehali:2021nsx}. Specifically, the effective field theory approach allows to establish the existence of a minimum of $S_{\text{eff}}(u; \Delta,\tau)$ at $u=0$, which spontaneously  breaks the one-form symmetry $\mathbb{Z}_N$ of the 4d $\mathcal{N}=4$ SYM theory. The fact that $u=0$ spontaneously breaks $\mathbb{Z}_N$ implies the existence of exactly $N-1$ additional local minima which contribute equally to the index, hence the $\log N$ correction to the logarithm of the SCI.

Summarizing, the logarithmic correction to the logarithm of the SCI, which we refer to as $\Delta \textrm{log}\, \mathcal{I}_{\text{CFT}_4}$, has been shown to be robust. In \cite{GonzalezLezcano:2020yeb} it was originally obtained using two different approaches to evaluate the index: the saddle-point approximation and the Bethe-Ansatz approach.  In the latter approach, the logarithmic term appears as the degeneracy of the Bethe-Ansatz solutions. The same logarithmic contribution was also shown to persist for a large class of ${\cal N}=1$ superconformal field theories. The form of the logarithmic correction was further confirmed  in  \cite{Aharony:2021zkr}, which  provides an interpretation  for certain exponentially suppressed terms. In  \cite{Amariti:2020jyx}, the logarithmic corrections were extended to other gauge groups and the results were shown to be compatible with the  $SU(N)$ analysis.

The black hole entropy is extracted from the SCI by implementing an inverse Laplace transformation, which yields the degeneracy of a state with given energy and charges. In the regime of large charges, we can reduce the inverse Laplace transformation to a Legendre transformation using the saddle point approximation. This is tantamount to changing from a grand-canonical ensemble to a microcanonical one. At the leading order in $N$, the two ways of approaching the entropy should be equivalent. However, when studying subleading structures we have to be more careful, since the very process of going from one ensemble to the other could modify the subleading corrections we are trying to probe. To be more specific, let us call $\Delta S_{\text{CFT}_4}$ the subleading logarithmic correction to the black hole entropy. Then we expect that in general $\Delta S_{\text{CFT}_4} =\Delta \textrm{log}\, \mathcal{I}_{\text{CFT}_4} + \text{(corrections from changing ensemble)}$. Let us now study more carefully the contribution coming from the change of ensemble.

\subsection{The Logarithmic Correction Associated to Changing Ensemble} \label{sec:4dGCMC}

We denote $\mathcal{I}_{\text{GC}}$ as the index computeded in the grand-canonical ensemble and $\mathcal{I}_{\text{MC}}$ as the index in the microcanonical ensemble,  i.e.,  the index for fixed values of the charges. We consider $D$ chemical potentials $\mu_{I}$ ($I=1, \cdots, D$) satisfying the constraint,
\begin{align}
\sum_{I=1}^D c_I\mu_I & =  n_0, \label{eq:constraint}
\end{align}
where $c_I =1$ for $\mu_I$ associated to electric charges and $c_I =-1$ for $\mu_I$ associated to angular momenta. We implement the inverse Laplace transform which takes us from the grand-canonical ensemble to the microcanonical ensemble
\begin{align}
\mathcal{I}_{\text{MC}} & = \int d^D \mu\, d \Lambda \exp \left\{ \log \mathcal{I}_{\text {GC}}- \sum_{I=1}^D  Q_I \mu_I - \Lambda\left(\sum_{I=1}^D c_I \mu_I - n_0\right)\right\},
\end{align}
where $\Lambda$ is the Lagrange multiplier associated to the constraint \eqref{eq:constraint}. Note that we have already considered the case of equal angular momenta when computing the index in the grand-canonical ensemble. Otherwise,  we would have also needed an additional Lagrange multiplier accounting for the constraint among rotations. We know the logarithmic corrections in the grand-canonical ensemble takes the form
\begin{align} \label{eq:GC result}
    \log \mathcal{I}_{\text{GC}} = \log \mathcal{I}_{\text{GC}}^{(\text{leading})} + \log N.
\end{align} 
Imposing \eqref{eq:GC result}, we find that the index in the microcanonical ensemble takes the form
\begin{align}
\mathcal{I}_{\text{MC}} & = N\int d^D \mu\, d \Lambda \exp \left[ \log \mathcal{I}_{\text{GC}}^{(\text{leading})}- \sum_{I=1}^D  Q_I \mu_I - \Lambda\left(\sum_{I=1}^D c_I \mu_I - n_0\right)\right]. \label{eq:Laplace}
\end{align}
We are now ready to implement the saddle point method, keeping the subleading logarithmic corrections associated to the one-loop determinant. The saddle point equations are given as
\begin{align}
\begin{split}
    \frac{\partial }{\partial \mu_{I}}\left[ \log \mathcal{I}_{\text{GC}}^{(\text{leading})}- \sum_{I=1}^D  Q_I \mu_I - \Lambda\left(\sum_{I=1}^D c_I \mu_I - n_0\right)\right] &=0,\\
    \frac{\partial }{\partial \Lambda}\left[ \log \mathcal{I}_{\text{GC}}^{(\text{leading})}- \sum_{I=1}^D  Q_I \mu_I - \Lambda\left(\sum_{I=1}^D c_I \mu_I - n_0\right)\right] & = 0,
\end{split}
\end{align}
which leads to
\begin{align}
    \begin{split}
    \frac{\partial \log \mathcal{I}_{\text{GC}}^{(\text{leading})}}{\partial \mu_I} & = Q_I + c_I \Lambda , \label{eq:saddle}\\
    \sum_{I=1}^D c_I\mu_I  =  n_0.
    \end{split}
\end{align}
A very important property of $\log \mathcal{I}_{\text{GC}}^{(\text{leading})}$ is its homogeneity of degree one in the chemical potentials. This implies the following crucial relation
\begin{align} \label{eq: homogeneity of I}
\log \mathcal{I}_{\text{GC}}^{(\text{leading})} & = \sum_{I=1}^D  \mu_I \frac{\partial \log \mathcal{I}_{\text{GC}}^{(\text{leading})}}{\partial \mu_I}.
\end{align}
Evaluating at the saddle point values, we obtain
\begin{align}
\log \mathcal{I}_{\text{GC}}^{\star(\text{leading})} & = \sum_{I=1}^D  \mu_I^{\star} \left(Q_I + c_I \Lambda\right),
\end{align}
such that the saddle point imposed on \eqref{eq:Laplace} yields
\begin{align}
\begin{split}
\mathcal{I}_{\text{MC}} & \approx N \exp \left\{\log \mathcal{I}_{\text{GC}}^{\star(\text{leading})} - \sum_{I=1}^D  Q_I \mu^{\star}_I- \Lambda\left(\sum_{I=1}^D c_I \mu^{\star}_I - n_0\right) -\frac{1}{2} \log \det \left(H\right)\right\}\\
& =  N \exp \left\{ \sum_{I=1}^D  \mu_I^{\star} \left(Q_I + c_I \Lambda\right) - \sum_{I=1}^D  Q_I \mu^{\star}_I- \Lambda\left(\sum_{I=1}^D c_I \mu^{\star}_I - n_0\right) -\frac{1}{2} \log \det \left(H\right)\right\}\\
& = N e^{n_0 \Lambda -\frac{1}{2} \log \det \left(H\right)}.
\end{split}
\end{align}
The Hessian $H$ has the form
\begin{align}
 H &  = \begin{pmatrix}
\frac{\partial^2 \log \mathcal{I}_{\text{GC}}^{(\text{leading})}}{\partial \mu_1^2} && \cdots && \frac{\partial^2 \log \mathcal{I}_{\text{GC}}^{(\text{leading})}}{\partial \mu_1 \partial \mu_D}  &&\frac{\partial^2 \log \mathcal{I}_{\text{GC}}^{(\text{leading})}}{\partial \mu_1 \partial \Lambda} \\
. && . && . && . \\
. && . && . && . \\
\frac{\partial^2 \log \mathcal{I}_{\text{GC}}^{(\text{leading})}}{\partial \mu_D \partial \mu_1} && \cdots && \frac{\partial^2 \log \mathcal{I}_{\text{GC}}^{(\text{leading})}}{\partial \mu_D^2} && \frac{\partial^2 \log \mathcal{I}_{\text{GC}}^{(\text{leading})}}{\partial \mu_D \partial \Lambda}  \\
\frac{\partial^2 \log \mathcal{I}_{\text{GC}}^{(\text{leading})}}{ \partial \Lambda\partial \mu_1}  && \cdots && \frac{\partial^2 \log \mathcal{I}_{\text{GC}}^{(\text{leading})}}{\partial \Lambda \partial \mu_D}  && \frac{\partial^2 \log \mathcal{I}_{\text{GC}}^{(\text{leading})}}{\partial \Lambda^2 }  \\
\end{pmatrix}.
\end{align}

Since $\log \mathcal{I}_{\text{GC}}^{(\text{leading})}$ is a homogeneous function of degree one, the chemical potentials can appear either in the numerator or the denominator in a way that the second derivative terms appearing along the diagonal of $H$ vanish when $\mu_I$ appears in the denominator. To keep track of this we define a list of numbers $\{\delta_{1} ,\cdots, \delta_{D}\}$ such that $\delta_I$ vanishes when $\mu_I$ is in the numerator of $\log \mathcal{I}_{\text{GC}}^{(\text{leading})}$ and it is equal to one otherwise.  This implies the following scaling of $H$
\begin{align}
\det H & \sim \det \begin{pmatrix}
\mathcal{O}(N^2)\delta_1 && \cdots && \mathcal{O}(N^2)  && c_1\\
. && . && . && . \\
. && . && . && . \\
\mathcal{O}(N^2) && \cdots && \mathcal{O}(N^2)\delta_D && c_D \\
c_1 && \cdots && c_D && 0 \\
\end{pmatrix} \sim \mathcal{O}(N^{2(D-1)}).
\end{align}
Defining $D=d+1$, where $d$ is the number of independent chemical potentials, the index computed in the microcanonical ensemble up to logarithmic corrections takes the form
\begin{align}\label{eq:logIMC from CFT4}
\log \mathcal{I}_{\text{MC}} \approx n_0 \Lambda +(1 - d) \log N.
\end{align}
From \eqref{eq:logIMC from CFT4} it is clear that from the CFT$_4$ point of view the origin of the logarithmic correction is two-fold: a part $\textrm{log}\, N$ from the grand-canonical ensemble and another part $-d\, \textrm{log}\, N$ from changing the grand-canonical ensemble to the microcanonical ensemble.  Since the chemical potentials are constrained by having equal angular momenta as well as the BPS condition, we have $d=3$ and therefore the logarithmic correction in the microcanonical ensemble is
\begin{align} \label{eq:log correction in MC CFT4}
    \Delta S_{\text{CFT}_4} &= -2\log N.
\end{align}
We expect this 4-dimensional result to match with the subleading correction coming from the 2-dimensional Cardy formula.

\subsection{Black Hole, Its Entropy and Near-Horizon Limit} 
The non-extremal asymptotically AdS$_5$ black hole background was found in \cite{Chong:2005hr}. In the Boyer-Lindquist coordinates $x^\mu = (t,\, r,\, \theta,\, \phi,\, \psi)$, the metric and the gauge field are given by \footnote{For simplicity, we consider the black hole with equal angular momenta $J_1=J_2$ and equal electric charges $Q_1 = Q_2 = Q_3$.}
\begin{align}
  ds^2 & = - \frac{\left[(1 + g^2 r^2) \rho^2 dt + 2 q \nu \right]\, dt}{\Xi \rho^2} + \frac{2 q }{\rho^2\Xi} \nu^2+ \frac{f}{\rho^4\Xi^2} \left(dt - \nu \right)^2   \nonumber\\
  {} & \quad + \frac{\rho^2 dr^2}{\Delta_r}+  \frac{\rho^2}{\Xi} \left( d\theta^2+\,\textrm{sin}^2 \theta\, d\phi^2 + \, \textrm{cos}^2 \theta\, d\psi^2\right)\, ,\label{eq:AdS5Metric}\\
  A & = \frac{\sqrt{3}\, q}{\rho^2\Xi} \left( dt - \nu\right)\, ,
\end{align}
where
\begin{align}
\begin{split}
  \nu & \equiv a\,(\, \textrm{sin}^2 \theta\, d\phi + \textrm{cos}^2 \theta\, d\psi)\, , \quad 
 \Xi  \equiv 1 - a^2 g^2\, ,\\
   \Delta_r & \equiv \frac{(r^2 + a^2)^2(1 + g^2 r^2) + q^2 + 2 a^2 q}{r^2} - 2 m\, , \label{eq:Delta-r}\\
  \rho^2 & \equiv r^2 +a^2 ,    \quad f \equiv 2 m \rho^2 - q^2 + 2 a^2 q g^2 \rho^2\, .
\end{split}
\end{align}
These black hole solutions are characterized by three independent parameters $(m,a, q)$, and $g$ is the inverse radius of AdS$_5$.  

We are ultimately interested in exploring the black hole solution for the parameter space satisfying supersymmetry and extremality, i.e. BPS. The supersymmetric limit corresponds to
\begin{align} \label{condition 1}
	q = \frac{m}{1+2ag}\, . 
\end{align}
However, this is not enough to ensure physical solutions and therefore we must also consider an additional constraint to prevent naked closed timelike curves, which in the BPS limit takes the form
\begin{align} \label{condition 2}
	m = \frac{2a(1+ag)^2(1+2ag)}{g}\, .
\end{align}
Extremality occurs when the inner horizon and the outer horizon coincide, which for our solution gives the double root
\begin{align}
	r_0^2 = \frac{a(2+ag)}{g}\, .
\end{align}

The macroscopic Bekenstein-Hawking entropy for the supersymmetric black hole, computed as a quarter of the area of the horizon  (in units of  $G_N = 1$), is
\be
S_{\text{BH}}=\frac{\pi^2 a^{3/2}\sqrt{2+ag}}{g^{3/2} (1-ag)^2} = 2\pi \sqrt{\frac{3Q^2}{g^2}-\frac{\pi}{2g^3}J}\, , \label{Eq:S_BH}
\ee
where  we have written it explicitly in terms of the electric charge, $Q$, and the angular momentum, $J$. The remarkable achievement of \cite{Cabo-Bizet:2018ehj, Choi:2018hmj, Benini:2018ywd} was to obtain this expression for the black hole entropy as the Legendre transform of the  leading $N^2$-part of the SCI \eqref{eq:SCI}, thus providing it with a microscopic explanation.

Given that the AdS/CFT correspondence geometrizes RG flow in the radial direction,  it is convenient to consider zooming into  a near-horizon region (IR), $r_0$, while assuming a  co-rotating frame:
\be\label{eq:AdS5scaling}
  r \to r_0 + \lambda\, \widetilde{r}\, ,\quad t \to \frac{\widetilde{t}}{\lambda}\, ,\quad \phi \to \widetilde{\phi} + g \frac{\widetilde{t}}{\lambda}\, ,\quad \psi \to \widetilde{\psi} + g \frac{\widetilde{t}}{\lambda}\, ,
\ee
where we have also imposed both \eqref{condition 1} and \eqref{condition 2}.  Taking $\lambda \to 0$ brings us to a  near-horizon region of the AdS$_5$ BPS black hole:
\begin{align}
  ds^2 & = \alpha_1 \left[- \widetilde{r}^2\, d\tau^2 + \frac{d\widetilde{r}^2}{\widetilde{r}^2}\right] + \Lambda_1(\theta) \left[d\widetilde{\phi} + \alpha_2 \, \widetilde{r}\, d\tau \right]^2 \nonumber\\
  {} & \quad + \Lambda_2(\theta)\left[d\widetilde{\psi} +\beta_1(\theta) d\widetilde{\phi} + \beta_2 (\theta)\, \widetilde{r}\, d\tau \right]^2 + \alpha_3\, d\theta^2\, ,\label{Eq:NH}
\end{align}
where
\begin{align}
\begin{split}
	\alpha_{1} &= \frac{a}{2 g(1+5 a g)}\, ,
	\\
	\alpha_{2} &=\frac{3a(1-ag)}{2(1+5ag)\sqrt{a\left(a+\frac{2}{g}\right)}}\, ,
	\\
	\alpha_{3} &= \frac{2a}{g(1- a g)}\, ,
	\\
	\Lambda_{1}(\theta) &=\frac{4 a(2+a g) \sin ^{2} \theta}{g(1-a g)(4-a g+3 a g \cos (2\theta))}\, ,
	\\
	\Lambda_{2} (\theta) &=\frac{a(4-a g+3 a g \cos (2 \theta)) \cos ^{2} \theta}{2 g(1-ag)^{2}}\, ,
	\\
	\beta_{1}(\theta)&= \frac{6 a g \sin ^{2} \theta}{4-a g+3 a g \cos (2 \theta)}\, ,
	\\
	\beta_{2}(\theta) &= \frac{3 g(1-a g) \sqrt{a\left(a+\frac{2}{g}\right)}}{(1+5ag)(4-ag+3ag\cos (2\theta))}\, .
\end{split}
\end{align}

 It is in the near-horizon limit at extremality where we find that  the near-horizon geometry is locally a $U(1)^2$-bundle over AdS$_2$. The asymptotic symmetries of this space can be studied via the Kerr/CFT correspondence, which  associates to  each $U(1)$-fiber in  \eqref{Eq:NH} a central charge and an effective temperature in the  CFT$_2$.  We can apply the Kerr/CFT correspondence to either $U(1)$, and the results of the black hole entropy from the Cardy formula are the same \cite{Chow:2008dp, David:2020ems}.

\subsection{Kerr/CFT Correspondence and Charged Cardy Formula}

Let us briefly review the Cardy formula which determines the degeneracy of states in a CFT$_2$. We are interested in its application up to and including the  logarithmic corrections to the degeneracy of states, with constraints among the charges and chemical potentials. We consider the partition function of a CFT$_2$ with  $n$ global $U(1)$ symmetries expressed in the grand-canonical ensemble
\be
  Z (\tau, \bar{\tau}, \vec{\mu}) = \textrm{Tr}\, e^{2 \pi i \tau L_0 - 2 \pi i \bar{\tau} \bar{L}_0 + 2 \pi i \mu_i P^i}\, , \label{eq:Zoriginal}
\ee
where $P^i$ are the conserved charges of the global $U(1)$'s, and $\mu_i$ are the corresponding chemical potentials. One particular property of the CFT$_2$ with conserved currents is that under modular transformations
\begin{align}
\tau \rightarrow \tau^{\prime}=\frac{a \tau+b}{c \tau+d}\, , \quad \mu_i \rightarrow \mu_i^{\prime}=\frac{\mu_i}{c \tau+d}\, , \hspace{3mm} i=1, \cdots, n,
\end{align}
the partition function transforms as (in a special choice of normalization)
\begin{align}
Z\left(\tau^{\prime}, \vec{\mu}^{\prime}\right)=e^{- 2\pi i \left(\frac{c \mu^{2}}{c \tau+d}\right)} Z(\tau, \vec{\mu})\, .
\end{align}
Therefore, the modular invariance of the partition function requires 
\be\label{Eq:modular-inv}
  Z (\tau, \bar{\tau}, \vec{\mu}) = e^{- \frac{2 \pi i \mu^2}{\tau}}\, Z \left(- \frac{1}{\tau},\, - \frac{1}{\bar{\tau}},\, \frac{\vec{\mu}}{\tau} \right)\, ,
\ee
where $\mu^2 \equiv \mu_i \mu_j k^{ij}$ with $k^{ij}$ denoting the matrix of the Kac-Moody levels of the $U(1)$ currents. The modular invariance \eqref{Eq:modular-inv}  implies that for small $\tau$
\be
  Z (\tau, \bar{\tau}, \vec{\mu}) \approx e^{- \frac{2 \pi i \mu^2}{\tau}}\, e^{- \frac{2 \pi i E_L^v}{\tau} + \frac{2 \pi i E_R^v}{\bar{\tau}} + \frac{2 \pi i \mu_i p_v^i}{\tau}}\, , \label{eq:Zgc}
\ee
where $E^v_{L}$, $E^v_R$ and $p_v^i$ are the lowest eigenvalues of $L_0$, $\bar{L}_0$ and $P^i$ respectively. Moreover, we take $E^v_{L}$, $E^v_R$ to be negative,  and $p_v^i =0$, corresponding to an electrically neutral vacuum.  Note that \eqref{eq:Zgc} is the grand-canonical partition function and does not contain logarithmic corrections, as opposite to the analogous quantity \eqref{eq:SCI} in CFT$_4$. This is already an indication that the Kerr/CFT computation is probing a different object in the grand-canonical ensemble. Nevertheless, Kerr/CFT still encodes information about the actual degeneracy of states by transforming to the microcanonical ensemble, i.e., the black hole sector with fixed charges and angular momentum.

Let us take a moment to understand the charges $p_{i}$ of the theory,  which include the angular momenta $p_{1}, p_{2}$ and the electric charges $p_{3},p_{4},p_{5}$, originally coming from the AdS$_5$ black hole solution. Particularly, for the 5d BPS black hole of interest, such charges obey a linear constraint of the generic form
\begin{align}
   \sum_{i=1}^5 b^{i} p_i & = M, \label{eq:const}
\end{align}
where $b^{i}$ are some constant coefficients and $M$ is related to the mass of the black hole.  Therefore,   \eqref{eq:const} implements the BPS conditions \eqref{condition 1} and \eqref{condition 2}.  Since we are considering $p^i \sim \mathcal{O}(N^0)$,  it can be seen from \eqref{eq:const} that $M \sim N^{0}$.

In the grand-canonical ensemble, we fix chemical potentials and admit all values of charges.  We consider a linear constraint among chemical potentials
\begin{align} \label{eq:constraint 2d 1}
    \sum_{i=1}^5 s^i \mu_i = C\, ,
\end{align}
where $C$ is a constant of the order $\mathcal{O} (N^0)$.  The constraint \eqref{eq:constraint} is a special case of \eqref{eq:constraint 2d 1}.  As we are going to see, this leads to the result that in terms of the scaling of $N$, $s^i \sim k_{ii}^{-1}$.  Moreover, in order to compare to the CFT$_{4}$ with equal angular momenta,  we consider an additional constraint of the form
\begin{align} \label{eq:constraint 2d 2}
    \sum_{i=1}^{2} \alpha^i \mu_i = 0,
\end{align}
of which the constraint $\mu_1 = \mu_2$ is a special case.  To clarify how we use these constraints to derive the logarithmic corrections, we carefully change to the microcanonical ensemble by integrating over chemical potentials while respecting the constraints \eqref{eq:constraint 2d 1} and \eqref{eq:constraint 2d 2}. The density of states $\rho(\tau, \bar{\tau}, \vec{\mu})$ can be expressed as the inverse Fourier transform of $Z(\tau, \bar{\tau}, \vec{\mu})$
\begin{align}
\begin{split}
  &\rho (E_L, E_R, \vec{p}) \\& =  \int d \tau\, d \Bar{\tau}\, d^n \mu\, d \lambda_1\, d\lambda_2\, \exp\left[2 \pi i S(\mu, \tau, \bar{\tau}) + 2 \pi i \lambda_{1} \left(\sum_{i=1}^5 s^i \mu_i-C\right) + 2 \pi i \lambda_{2}\sum_{i=1}^{2} \alpha^i \mu_i
  \right], \label{eq:rho}
\end{split}
  \\
  S(\mu, \tau, \bar{\tau})   & = - \frac{\mu^2}{\tau} - \frac{E^{v}_L}{\tau} +\frac{E^v_{R}}{\bar{\tau}} +\frac{\mu_i p_v^i}{\tau} - \tau E_L + \Bar{\tau}E_R - \mu_i p^i\, , \label{eq: S in 2d}
\end{align}
where $E_L$, $E_R$ and $p^i$ are the eigenvalues of $L_0$, $\bar{L}_0$ and $P^i$, respectively, and $n$ denotes the number of independent chemical potentials. Before we proceed, we take a moment to discuss the scaling of the various expressions and parameters involved. This is a crucial step in understanding which terms contribute to the subleading corrections of the entropy. The modular parameters are order-1 parameters: $\tau, \bar \tau \sim \mathcal{O}(N^{0})$. Similarly, we take $p^{i} \sim N^{0}$ and $C \sim N^{0}$. From \eqref{eq: S in 2d}, we find that $\mu^{2} \sim \mu_{i}\, p^{i}\, \tau$, which solving for the scaling of $\mu_{i}$ gives
\begin{align}
    \mu_{i} \sim p^{j}k_{ij}\tau \quad \Rightarrow \quad \mu_{i} \sim \sum_{j}p^{j} k_{ij} \sim (s^{i})^{-1},    
\end{align}
where we have made the summation over the indices explicit, to make it clear that the highest order in the summation should be the scaling of $\mu_{i}$ and $s^{i}$.
Likewise, $E_{L} \sim E_{R} \sim E_{L}^{v} \sim E_{R}^{v} \sim \mu^2 \sim \sum_{i,j}p^{i}p^{j}k_{ij}.$

Therefore, we have related the different parameters to the matrix $k_{ij}$, where the scaling can be found via the Kac-Moody levels. There are two types of levels that we are interested in. The Kac-Moody level from the $SU(2)$ rotation,  i.e. $k^{11}$ or $k^{22}$,  is proportional to the central charge $c$ \cite{Maldacena:1997ih}, which is of the order of Newton's constant $G^{-1} \sim N^{2}$. The Kac-Moody levels from $U(1)$ gauge symmetries, i.e. $k^{ii}$ ($i > 2$), are proportional to $N^{-2}$ \cite{Pathak:2016vfc, Compere:2009dp}. For the BPS AdS$_5$ black hole, the $N$-dependences of various factors are
\be \label{eq: scalings 1}
  E_{L}^{v}, E_{R}^{v} \sim N^{2}\, ,\quad 4 E_{L} - \mathcal{P}^2, E_{R} \sim N^{2}\, ,\quad k^{11}, k^{22} \sim N^{2}\, ,\quad k^{33}, k^{44}, k^{55} \sim N^{-2}\, ,
\ee
which implies that $k_{11}, k_{22} \sim N^{-2}$ and $k_{33}, k_{44}, k_{55} \sim N^{2}$.  Moreover,  this also implies that
\begin{align} \label{eq: scalings 2}
    s^{1} = s^{2} \sim N^{2}, \quad s^{3} = s^{4} = s^{5} \sim N^{-2}, 
\end{align}
or likewise $s_{1} = s_{2} \sim N^{0}$,  $s_{3} = s_{4} = s_{5} \sim N^{0}$.  Due to the definition of $\mu_{i}$ and $p_{i}$ as $\mu^2 = \mu_{i}\mu_{j}k^{ij}$ and $\mathcal{P}^2 \equiv p_{i}p_{j}k^{ij}$, we lower and raise the indices of $\mu_{i}, p_{i}$ and $s_{i}$ with $k_{ij}$.  However, for $\alpha_{i}$ we do not need to raise indices with $k_{ij}$,  as $\alpha_{1}$ and $\alpha_{2}$ are the net scalings due to that the right-hand side of \eqref{eq:constraint 2d 2} is zero.

With these scalings in mind, we can now proceed to compute the saddle point.  From \eqref{eq:rho} and \eqref{eq: S in 2d} let us define
\be
  \widetilde{S} (\mu,\, \tau,\, \bar{\tau}) \equiv S (\mu,\, \tau,\, \bar{\tau}) + \lambda_{1} \left(\sum_{i=1}^5 s^i \mu_i-C\right) + \lambda_{2}\sum_{i=1}^{2} \alpha^i \mu_i\, .
\ee
The equations for the fixed points have the form
\begin{subequations}
\begin{align}
     i = 1,\, 2: \quad \frac{\partial \widetilde{S}}{\partial \mu^i} & = - 2 \frac{k_{ij} \mu^j}{\tau} + \frac{(p_{v})_i}{\tau} - p_i + \lambda_{1} s_i +  \lambda_{2} \alpha_{i} =0\, , \label{eq:lagrange mu}
    \\
    i = 3,\, 4,\, 5: \quad \frac{\partial \widetilde{S}}{\partial \mu^i} & = - 2 \frac{k_{ij} \mu^j}{\tau} + \frac{(p_{v})_i}{\tau} - p_i +  \lambda_{1} s_i=0\, ,
    \\
    \frac{\partial \widetilde{S}}{\partial \tau} & = \frac{\mu^2}{\tau^2} + \frac{E^{v}_{L}}{\tau^2} - \frac{\mu_i p^{i}_{v}}{\tau^2} - E_{L} = 0\, ,
    \\
    \frac{\partial \widetilde{S}}{\partial \bar \tau} & = - \frac{E^v_{R}}{\bar \tau^2} + E_{R} = 0\, ,
    \\
\frac{\partial \widetilde{S}}{\partial \lambda_{1}} & = \sum_{i=1}^5 s^i \mu_i - C = 0\, , \label{eq:lagrange lambda 1}
    \\
    \frac{\partial \widetilde{S}}{\partial \lambda_{2}} & =\sum_{i=1}^2 \alpha^i \mu_i = 0\, . \label{eq:lagrange lambda 2}
    \end{align}
\end{subequations}
We define the values of the saddle to be $(\mu_{i})_{0}$, $\tau_0$ and $\bar \tau_0$,  such that \eqref{eq:lagrange mu} gives
\begin{align}
    (\mu_{i})_{0} = \begin{cases} \frac{1}{2}k_{ij}\left(p^j_v - p^j \tau_0 +  \lambda_1 s^j \tau_0+  \lambda_2 \alpha^j \tau_{0} \right), &  \quad i=1,2\, ,
    \\
     \frac{1}{2}k_{ij}\left(p^j_v - p^j \tau_0 + \lambda_1 s^j \tau_0\right), & \quad i=3,4,5\, ,  \end{cases}
\end{align}
where we can redefine $p^i$ by shifting it as follows
\begin{align}\label{eq:def ptilde}
   \widetilde{p}^{i} \equiv \begin{cases} 
      p^i -\lambda_1 s^i -  \lambda_2 \alpha^i\, , & i=1,2\, , \\
      p^i -  \lambda_1 s^i\, , & i=3,4,5\, .\\
   \end{cases}
\end{align}
Therefore, we rewrite
\begin{align}
    (\mu_{i})_{0} & =\frac{1}{2}k_{ij}\left(p^j_v - \widetilde{p}^j \tau_0 \right)\, , \label{eq:mu0mod}
\end{align}
with $\tau_0$ satisfying
 \begin{align}
     \tau_0^2 E_L & = \mu_{0}^2 +E^v_L - (\mu_{i})_{0} p^i_v\, . \label{eq:tau02}
 \end{align}
Using \eqref{eq:mu0mod}, we find that
\begin{align} \label{eq:mu squared}
    \mu_{0}^2& = \frac{1}{4} \left[k_{ij} p_v^i p_v^j + \tau_0^2 k_{ij} \widetilde p^i \widetilde p^j\right]- \frac{1}{2}k_{ij}\widetilde p^i p^j_{v}\tau_0  = \frac{1}{4}k_{ij}\tau_0^2\widetilde p^i \widetilde p^j\, ,
\end{align}
where $k_{im}k^{il}=\delta_{m}^{l}$,  and in the second equality we have assumed that the vacuum is electrically neutral, i.e.  $p_v^i =0$. Inserting  \eqref{eq:mu squared} in \eqref{eq:tau02},  we obtain
\begin{align}
    \tau_0 & = \pm  i\sqrt{\frac{4(-E^v_L)}{4E_L-k_{ij}p^i p^j}}\, .
\end{align}
The saddle point for $\Bar{\tau}$ trivially is $\Bar{\tau}_0= \pm i \sqrt{\frac{-E^v_R}{E_R}}$.  Consequently,  $(\mu_{i})_{0}$ given by \eqref{eq:mu0mod} now has the form
\begin{align}
    (\mu_{i})_{0}& =- \frac{1}{2} k_{ij} \widetilde p^{j}\tau_{0}=\mp i k_{ij}\widetilde p^j\sqrt{\frac{-E^v_L}{4E_L-k_{ij}\widetilde p^i \widetilde p^j}}\, .
\end{align}
Imposing \eqref{eq:lagrange lambda 2}, we find 
\begin{align}
    \sum_{i=1}^{2}\alpha^{i}(\mu_{i})_{0}=- k_{ij}\alpha^i\widetilde{p}^j\sqrt{\frac{E^v_L}{4E_L - k_{ij}\widetilde{p}^i\widetilde{p}^j}} = 0 \quad \Rightarrow \quad \alpha^{1}k_{11}\widetilde p^{1}+\alpha^{2}k_{22}\widetilde p^{2}=0\, .
\end{align}
Choosing the normalization of the Kac-Moody levels such that $k_{11} = k_{22}$, we can solve for the Lagrange multiplier $\lambda_{2}$
\begin{align}
\begin{split}\label{eq:intermediate}
    \alpha^{1}k_{11}\widetilde p^{1}+\alpha^{2}k_{22}\widetilde p^{2}& =0\, , \\
    \Rightarrow\quad k_{11} \left( \alpha^1 (p^1 - \lambda_1 s^1 - \lambda_2 \alpha^1) +\alpha^2(p^2- \lambda_1 s^2 - \lambda_2 \alpha^1)\right) & = 0\, , \\
    \Rightarrow\quad \alpha^1 p^1 + \alpha^2 p^2 - \lambda_1 (\alpha^1 s^2 + \alpha^2 s^2)  -\lambda_2( (\alpha^1)^2 + (\alpha^2)^2) &= 0\, .
    \end{split}
\end{align}
We now set $\alpha^{1} = - \alpha^{2}$,  since both $p^{1}$ and $p^{2}$ should have the same scaling.  Therefore, we obtain
\begin{align} \label{eq: scalings 3}
    \alpha_{1},\alpha_{2} \sim 1,
\end{align}
and
\be
  \alpha^{1} \sim \alpha^{2} \sim 1\, ,
\ee
where we do not need to raise indices with $k_{ij}$ here,  as $\alpha_{1}$ and $\alpha_{2}$ are the net scalings due to that the right-hand side of \eqref{eq:constraint 2d 2} is zero.  Moreover, $s^{1} = s^{2}$ as they correspond to the equal angular momenta. We then find from \eqref{eq:intermediate} that
\begin{align}
    \begin{split}
        \alpha^1(p^1 - p^2) & = 2 (\alpha^1) ^2 \lambda_2\, , 
        \\
        \Rightarrow\quad \lambda_2 & = \frac{p^1- p^2}{2 \alpha^1}\, ,
    \end{split}
\end{align}
which vanish for $p^1 = p^2$. This implies that $\lambda_{2}$ does not affect the logarithmic corrections to the entropy for the case of equal angular momenta,  as its contribution to the determinant of the Hessian matrix is of the order $\mathcal{O} (N^0)$.  We are also interested in the scaling of $\lambda_{1}$.  From \eqref{eq:lagrange lambda 1}, we find
\begin{align}\label{eq:intermediate2}
    \sum_{i}^{5}s^{i}(\mu_{i})_{0}=- k_{ij}s^i\widetilde{p}^j\sqrt{\frac{E^v_L}{4E_L - k_{ij}\widetilde{p}^i\widetilde{p}^j}}& = C\, .
\end{align}
As we expect the scaling to remain the same for any arbitrary values of the charges, we consider a special case $p^i = p_v^i =0$ and find from \eqref{eq:def ptilde} and \eqref{eq:intermediate2} that
\begin{align}
\begin{split}
  {} & \lambda_1\sum_{i=1}^5 k_{ii} s^i s^i = C \sqrt{\frac{4 E_L - \lambda_1^2\sum_{i=1}^5k_{ii}s^is^i}{E_L^v}}\, ,\\
   \Rightarrow\quad & \lambda_1^2 \left(\sum_{i=1}^5 k_{ii} s^i s^i \right)^2 = C^2 \left(\frac{4 E_L}{E_L^v} - \frac{\lambda_1^2\sum_{i=1}^5k_{ii}s^is^i}{E_L^v}\right)\, ,\\
   \Rightarrow\quad & \lambda_1^2\left[\left(\sum_{i=1}^5 k_{ii} s^i s^i\right)^2 + C^2\frac{\sum_{i=1}^5k_{ii}s^is^i}{E_L^v}\right] = 4 C^2 \frac{E_L}{E_L^v}\, .
    \end{split}
\end{align}
Let us now discuss the scalings of each of these terms. Given \eqref{eq: scalings 1} and \eqref{eq: scalings 2}, we have at the leading order $E_{L} \sim E_{L}^{v} \sim N^2$,  $C \sim N^{0}$ and $k_{ii} s^i s^i \sim N^2$ and therefore
\begin{align}
    \lambda_1 & \sim N^{-2}.
\end{align}

The leading order value of the degeneracy is obtained by evaluating the action at the saddle point values,  which gives
\begin{align} \label{eq: log rho with p tilde}
\log \rho_0 & = \sqrt{E^v_{L} \left(4 E_L - k_{ij}\widetilde{p}^i\widetilde{p}^j\right)} +2\sqrt{E_R E^v_R}\, .
\end{align}
We would like to comment on the scaling with respect to $N$ in \eqref{eq: log rho with p tilde}.  At the leading order, $\lambda_1 s^i \sim \mathcal{O}(N^{-2})$ which implies that $\mathcal{O}(\widetilde{p}^i) \sim \mathcal{O}(p^i)$. Therefore, \eqref{eq: log rho with p tilde} coincides with the leading order of the degeneracy with $\widetilde p^{i}$ replaced by $p^{i}$. This is important as we can see that the constraint we imposed only affects the subleading order of the entropy.

Note that we have more than one saddle points,  namely one for each choice of signs in the values of $\tau_0, \Bar{\tau}_0, (\mu_{i})_{0}$. However, one saddle dominates over the others as $S(\mu,\tau,\bar\tau)$ is exponentially suppressed. To see this explicitly, let us take
\be\label{eq:saddle}
  \tau_0 = i \epsilon_{\tau} \sqrt{\frac{4(-E^v_L)}{4E_L-k_{ij}p^i p^j}}\, ,\quad\quad \bar{\tau}_0 =  i \epsilon_{\bar \tau} \sqrt{\frac{-E^v_R}{E_R}}\, , \quad\quad \mu_{i,0}  = - i \epsilon_{\tau} k_{ij}\, p^j\, \sqrt{\frac{-E^v_L}{4E_L-k_{ij}p^i p^j}}\, ,
\ee
where $\epsilon_{\tau}$ and $\epsilon_{\bar \tau}$ take on values of $\pm 1$. Then,  under the constraints imposed by the Lagrange multipliers,  the density of states \eqref{eq:rho} can be approximated by the saddle points
\begin{align}
\begin{split}
\rho_0 &  =\sum_{\substack{\epsilon_{\tau}, \epsilon_{\bar \tau} = \pm 1}} \exp \left(2 \pi i S(\mu,\tau,\bar \tau)\right) \\
& =  \displaystyle \sum_{\substack{\epsilon_{\tau}, \epsilon_{\bar \tau} = \pm 1}}\exp\left\{2 \pi i\left[-\frac{i \epsilon_{\tau}}{2}\frac{\mathcal{P}^2\sqrt{-E_L^{v}}}{\sqrt{4E_{L}-\mathcal{P}^2}} -i \epsilon_{\tau}(-E_{L}^{v}) \sqrt{\frac{4E_{L}-\mathcal{P}^2}{4(-E_{L}^{v})}} + i \epsilon_{\bar \tau} (-E_{R}^{v}) \sqrt{\frac{E_{R}}{-E_{R}^{v}}} \right. \right. \\ & \qquad \qquad \qquad \qquad \qquad \left. \left. - i\epsilon_{\tau}E_{L} \sqrt{\frac{-4E_{L}^{v}}{4E_{L}-\mathcal{P}^2}}+ i\epsilon_{\bar \tau} E_{R} \sqrt{\frac{-E_{R}^{v}}{E_{R}}} -(-i\epsilon_{\tau}) \mathcal{P}^2\sqrt{\frac{-E_{L}^{v}}{4E_{L} - \mathcal{P}^2}}\right] \right\} \\
& =\sum_{\substack{\epsilon_{\tau}, \epsilon_{\bar \tau} = \pm 1}} \exp\left[ 2\pi \left( \epsilon_{\tau} \sqrt{-E_{L}^{v} (4E_{L} - \mathcal{P}^2)} - \epsilon_{\bar \tau} \sqrt{-4E_{R}^{v} E_{R}}\right)\right]\, . \label{eq:saddlescorr}
\end{split}
\end{align}
If we now select the combination of $\epsilon_{\tau} = 1$ and $\epsilon_{\Bar{\tau}}=-1$,  which maximizes the exponent in \eqref{eq:saddlescorr},  we can write
\begin{align}
\begin{split}
\rho_0 &  \sim  \exp\left[ 2\pi \left(\displaystyle  \sqrt{-E_{L}^{v} (4E_{L} - \mathcal{P}^2)} + \sqrt{-4E_{R}^{v} E_{R}}\right)\right]+ \dots\, ,
\end{split}
\end{align}
where the dots denote the exponentially suppressed terms of subleading non-logarithmic order. To summarize, given the behavior of $Z$, the degeneracy $\rho$ can be determined using the saddle-point approximation with the dominant saddle at 
\be\label{eq:saddle}
  \tau_0 = \sqrt{\frac{4 E_L^v}{4 E_L - \mathcal{P}^2}}\, ,\quad \bar{\tau}_0 = - \sqrt{\frac{E_R^v}{E_R}}\, ,\quad \mu_{i,0} = - k_{ij}\, p^j\, \sqrt{\frac{E_L^v}{4 E_L - \mathcal{P}^2}}\, ,
\ee
where $k_{ij}$ is the inverse matrix of $k^{ij}$, and $\mathcal{P}^2 \equiv p^i p^j k_{ij}$. Note that the saddle-point values $\tau_0$ and $(\mu_{i})_{0}$ are parametrically small, as $\mathcal{P}^2 \gg |E_{L}|$, which is reminiscent of the 4d Cardy limit originally used in  \cite{Cabo-Bizet:2018ehj, Choi:2018hmj} and recently clarified in \cite{Cassani:2021fyv, ArabiArdehali:2021nsx}. Moreover, the Cardy limit in 2d, which assumes that the levels of the theory is much larger than the Casimir energy, is compatible with the Cardy limit in 4d, which focuses on small chemical potentials and large charges, as they both address the high energy states of the theory and in our particular case address the entropy of extremal black holes.

At the saddle \eqref{eq:saddle}, the density of states $\rho$ reaches its extremum $\rho_0$, and the corresponding entropy is
\be\label{eq:S from saddle}
  S (\mu_{0}, \tau_0, \bar \tau_0) = \textrm{log}\, \rho_0 \approx 2 \pi \sqrt{- E_L^v (4 E_L - \mathcal{P}^2)} + 2 \pi \sqrt{- E_R^v (4 E_R)}\, .
\ee
This expression is also called the charged Cardy formula in \cite{Hosseini:2020vgl}, which implies a microcanonical ensemble of black hole microstates. If we apply $E_L^v = E_R^v = - c / 24$, and define the temperatures $T_{L, R}$ through
\be
  E_L - \frac{\mathcal{P}^2}{4} = \frac{\pi^2}{6} c T_L^2\, ,\quad E_R = \frac{\pi^2}{6} c T_R^2\, ,
\ee
we can rewrite the entropy \eqref{eq:S from saddle} as
\be\label{eq:S from Cardy}
  S = \frac{\pi^2}{3} c T_L + \frac{\pi^2}{3} c T_R\, ,
\ee
where $T_R$ is proportional to the physical Hawking temperature $T_H$. This formula coincides,  at the leading order,  with the canonical ensemble version of the charged Cardy formula, and has been successfully used in a variety of cases \cite{Guica:2008mu, Castro:2010fd, Haco:2018ske, Haco:2019ggi, Perry:2020ndy}.  However, we emphasize that obtaining \eqref{eq:S from Cardy} did not involve a change of ensemble,  as we merely re-identified certain combinations.

From the near-horizon CFT$_2$ and the Kerr/CFT correspondence we know that for the BPS AdS$_5$ black hole
\begin{align}
  c_L & = \frac{9 \pi a^2}{G_N g (1 - a g) (1 + 5 a g)} = \frac{18 N^2 (a g)^2}{(1 - a g) (1 + 5 a g)}\, ,\\
  T_L & = \frac{1 + 5 a g}{3 a (1 - a g) \pi} \sqrt{a \left(a + \frac{2}{g} \right)}\, ,
\end{align}
where we have used the AdS$_5$/CFT$_4$ dictionary $ \frac{1}{2} N^2 = \frac{\pi}{4 G_N} \ell_5^3 = \frac{\pi}{4 G_N g^3}$. Note that both $c_L$ and $T_L$ are dimensionless. Consequently, the BPS AdS$_5$ black hole entropy at the leading order in $N$ is given by the Cardy formula
\begin{align}
  S_{\text{CFT}_2} & = \frac{\pi^2}{3} c_L T_L  = \frac{2 N^2 \pi (a g)^{3/2} \sqrt{ 2+ a g}}{(1 - a g)^2} \, .\label{eq:AdS5 BH S Cardy}
\end{align}
This near-horizon CFT$_2$ result matches the macroscopic Bekenstein-Hawking entropy of the black hole \eqref{Eq:S_BH}, as shown in \cite{Lu:2008jk, Chow:2008dp, David:2020ems}.

\subsection{Logarithmic Corrections  from  Near-Horizon CFT$_2$} \label{sec:2dGCMC}

To derive the logarithmic corrections to the black hole entropy from the near-horizon CFT$_2$,  we evaluate the Cardy formula beyond its leading saddle-point value by including its Gaussian correction. Namely,  we consider a  logarithmic correction $\Delta S_{\text{CFT}_2}$ obtained from expanding $\tau$, $\bar{\tau}$ and $\vec{\mu}$ to the quadratic order around the saddle point given by \eqref{eq:saddle}. The result is
\be
  \Delta S_{\text{CFT}_2} = - \frac{1}{2}\, \textrm{log}\, \frac{\textrm{det}\, \mathcal{A}}{(2 \pi)^{n+2}}\, ,
\ee
where $\mathcal{A}$ is the Hessian of the exponent in the integrand of \eqref{eq:rho} around the saddle point \eqref{eq:saddle}, and
has the form  $\mathcal{A}_{\mu \nu}=\frac{\partial^2 \widetilde{S}}{\partial x^{\mu}\partial x^{\nu}}$, where $x^{\mu} = \{\tau, \Bar{\tau}, \lambda,\mu^{i=1,\cdots,n}\}$, whose only non-trivial elements in the presence of constraints are
\begin{align}
\begin{split}
  \frac{\partial^2 \widetilde{S}}{\partial \tau \partial \mu^i} & =2 \frac{k_{ij}(\mu^{j})_0 }{\tau_0},
    \\
  \frac{\partial ^2 \widetilde{S}}{\partial \tau^2}  & = -\frac{2}{\tau_0^3}\left(k_{ij}(\mu^{i})_0(\mu^{j})_0 +E^v_{L} \right),
       \\
  \frac{\partial ^2 \widetilde{S}}{\partial \bar{\tau}^2} & = 2 \frac{E^v_R}{\Bar{\tau}_0^3},
   \\
  \frac{\partial^2 \widetilde{S}}{\partial \lambda_{1} \partial \mu^i} &= s_{i},\quad (i = 1,\, \cdots,\, 5)
   \\
  \frac{\partial^2 \widetilde{S}}{\partial \lambda_{2} \partial \mu^i} &= \alpha_{i},\quad (i = 1,  2)
    \\
  \frac{\partial^2 \widetilde{S}}{\partial \mu^i \partial \mu^j}& = -2\frac{k_{ij}}{\tau_0}.
\end{split}  
\end{align}
We see that $k_{11}$ and $k_{22}$ come from the $SU(2)$ rotation, which corresponds to the angular momenta, while $k^{ii}$ ($i > 2$) come from the $U(1)$ gauge symmetries.  At the subleading order the Hessian takes the form
\be\label{eq:detA}
  \textrm{det}\, \mathcal{A} = \frac{(2 \pi)^{n+2}}{16}\, (- E_L^v)^{- \frac{n+1}{2}}\, (4 E_L - \mathcal{P}^2)^{\frac{n+3}{2}}\, (- E_R^v)^{- \frac{1}{2}}\, (4 E_R)^{\frac{3}{2}}\, \textrm{det} (H)\, ,
\ee
where
\begin{align}
    \renewcommand{\arraystretch}{1.5}
   H \sim
    \left(\begin{array}{c|c}
    k_{ij} &  \mathcal{S}^{T}  \\ \hline
    \mathcal{S} &  0
    \end{array}\right),
\end{align}
and
\begin{align}
    \mathcal{S} &= \left(\begin{array}{ccccc}
    s_{1} &  s_{2} &  s_{3} &  s_{4} &  s_{5}  \\
     \alpha_{1}  & \alpha_{2}  & 0 & 0 & 0 
    \end{array}
    \right).
\end{align}
Note that this result is different than in \cite{Pathak:2016vfc}, as we have considered two linear constraints on the chemical potentials.

For supersymmetric extremal (BPS) black holes,  one of the Frolov-Thorne temperatures $T_R$ vanishes,  as it is proportional to the Hawking temperature,  and only the left sector contributes to the black hole entropy. Consequently, \eqref{eq:detA} for BPS black holes becomes
\be\label{eq:BPSdetA}
  \left(\textrm{det}\, \mathcal{A}\right)_{\text{BPS}} = \frac{(2 \pi)^{n+2}}{16}\, (- E_L^v)^{- \frac{n+1}{2}}\, (4 E_L - \mathcal{P}^2)^{\frac{n+3}{2}}\, \textrm{det} (H)\, ,
\ee
where
\be
  - E_L^v = \frac{c}{24}\, ,\quad 4 E_L - \mathcal{P}^2 = \frac{2 \pi^2}{3} c T_L^2\, .
\ee
With the scalings in \eqref{eq: scalings 1}, \eqref{eq: scalings 2} and \eqref{eq: scalings 3}, the Hessian takes on the $N$-dependence
\begin{align}
    \det H \sim N^{2},
\end{align}
such that
\be
  (\textrm{det}\, \mathcal{A})_{\textrm{AdS$_5$ Black Hole}} \sim (N^2)^{- \frac{n+1}{2}}\, (N^2)^{\frac{n+3}{2}}\, \left(N^2\right) = N^{4}\,.
\ee
Note that the result is independent of $n$ since the scaling of $E_{L}^{v}$ and $4E_{L}-\mathcal{P}^2$ are equal. Therefore, the logarithmic correction to the leading-order BPS AdS$_5$ black hole entropy \eqref{eq:AdS5 BH S Cardy} is
\be \label{eq: CFT2 result}
  \Delta S_{\text{CFT}_2} = - \frac{1}{2}\, \textrm{log}\, \frac{\textrm{det}\, \mathcal{A}}{(2 \pi)^{n+2}} = -2\, \textrm{log}\, N + \mathcal{O} (1)\, , 
\ee
which precisely agrees with $\Delta S_{\text{CFT}_4}$ in \eqref{eq:log correction in MC CFT4}.  However, we also observe that in contrast to the CFT$_4$ approach,  where the logarithmic correction originates from both the grand-canonical ensemble and the process of changing ensemble,  the logarithmic correction in the CFT$_2$ approach comes purely from changing the ensemble.  This is because the Kerr/CFT is intrinsically a near-horizon field theoretic approach,  and it does not probe the boundary of the full geometry,  where the values of the chemical potentials in the grand-canonical ensemble are fixed.  Hence,  in Kerr/CFT it is more natural to consider the entropy in the microcanonical ensemble,  and we only expect the results of the black hole entropy with logarithmic correction from the CFT$_4$ and the CFT$_2$ approaches match in this ensemble.

\section{AdS$_5$ Black Strings}\label{sec:AdS5BS}

\subsection{AdS$_5$ Black String Entropy from Boundary $\mathcal{N} = 4$ SYM} \label{sec:blackstring}

A rotating AdS$_5 $ black string  solution in gauged supergravity has been discussed in \cite{Hosseini:2019lkt, Hosseini:2020vgl}, where it was shown that its leading-order entropy  can  be obtained from the refined topologically twisted index of $\mathcal{N} = 4$ SYM on  $S^2\times T^2$.

The topologically twisted index of $\mathcal{N}=4$ SYM with gauge group $SU(N)$ is defined as the supersymmetric index of the theory on $T^2 \times S^2$ with a topological twist on $S^2$ \cite{Benini:2015noa,Hosseini:2016cyf}, its Hamiltonian interpretation being 
\be
  Z(p_a, \Delta_a)= (-1)^F e^{2\pi i \tau \,\{{\cal Q}, {\cal Q}^\dagger\}}e^{i \Delta_a J_a}\, .
\ee

 The topologically twisted index depends on a set of chemical potentials, $\Delta_{a}$,  for the generators of  flavor symmetries ($a=1,2,3$), a modular parameter of the torus $\tau$ and magnetic fluxes $p_a$. The topologically twisted index of ${\cal N}=4$ SYM with gauge group $SU(N)$ admits a presentation as an integral over the space of holonomies in the following way:
 \begin{align}
 \begin{split}
 Z(p_a, \Delta_a) & = \frac{1}{N!}\sum_{m} \oint_{\mathcal{C}}\prod_{\mu=1}^{N-1} \left(d u_{\mu} \eta(q)^2\right) \mathcal{Z}_{TT}(u, \Delta_a, \tau, p_a)\, , \label{eq:TTI}\\
 \mathcal{Z}_{TT}(u, \Delta_a, \tau, p_a)& =\prod_{i,j=1}^N\left[\frac{\theta_1\left(u_{ij}; \tau\right)}{i \eta(q)}\prod_{a=1}^3 \left(\frac{i \eta(q)}{\theta_1\left(u_{ij}+ \Delta_a; \tau\right)}\right)^{m_{ij}- p_a +1}\right]\, ,
 \end{split}
 \end{align}
where $\eta(q)$ is the Dedekind eta function that we define in Appendix \ref{app:SpecialFunctions}. We can evaluate \eqref{eq:TTI} as the sum over residues \cite{Hosseini:2016cyf} which takes the following explicit form: 
  \begin{align}
    Z(p_a, \Delta_a)&= \eta(q)^{2(N-1)}\sum_{\hat{u} \in \text{BA}}\prod_{i,j=1}^N\left[\prod_{a=1}^3 \left(\frac{\theta_1\left(u_{ij}; \tau\right)}{\theta_1\left(u_{ij}+ \Delta_a; \tau\right)}\right)^{1- p_a }\right]  H^{-1}(\hat{u},\Delta, \tau),\label{eq:TTIBA}
  \end{align}
  where, analogously to the SCI discussed in Sec.~\ref{sec:thesci}, BA stands for the set of solutions to the Bethe-Ansatz equations \eqref{BAEs}, and $ H(\hat{u},\Delta, \tau)$ is the Jacobian defined in \eqref{eq:index:BA}. The location of a set of such residues was found in \cite{Hong:2018viz} and have the form given by \eqref{eq:BAsolTT} labeled by $\{u_i\}$ with integers $\{m,n,r\}$.

In fact, the set of solutions found in \cite{Hong:2018viz} inspired the evaluation of the SCI carried in \cite{Benini:2018ywd}, where the $\{u_i\}$ are also organized according to equation \eqref{eq:BAsolTT}, however, in the large-$N$ limit, it was possible to argue that the configuration corresponding to $\{1,N,0\}$ was dominant. For fixed $\{m,n,r\}$, it is possible to count how many values of $\Bar{u}$ give non-equivalent contributions to topologically twisted index (by non-equivalent we mean, those which are not identified by periodicity $u\sim u+1$ or $u \sim u+\tau$). Once again, imposing the $SU(N)$ constraint we find that:
  \begin{align}
  \begin{split}
      \Bar{u} & = \frac{k}{N} -\frac{1}{2N}\left[n(m-1)+m (n-1)\left(\tau +\frac{r}{m}\right)\right]\, , \label{eq:TTubar}\\
      k& = 0,1,\cdots, N-1\, ,
      \end{split}
  \end{align}
  which reduces to \eqref{eq:ubar} for $\{m,n,r\}=\{1,N,0\}$. We then conclude that there is a degeneracy factor of $N$ for each $\{m,n,r\}$ configuration contributing to the topologically twisted index.
 To argue that there is no other contribution of the same order that spoils the value of the coefficient of $\log N$ would require a more detailed study of the large-$N$ behavior of the topologically twisted index, which has been studied recently in \cite{Hong:2021bzg} at the leading order in $N$. A systematic study of subleading corrections to the topologically twisted index still remains an open problem. It is, however, very tempting to conjecture that indeed, there is no contribution other than the one originated from degeneracy of Bethe-Ansatz solutions and, consequently, the coefficient of $\log N$ is $1$ also for the topologically twisted index in the grand-canonical ensemble.  
  
One can further refine the topologically twisted index by adding a rotation on $S^2$ \cite{Benini:2015noa}. This will modify the integral expression \eqref{eq:TTI} through the appropriate fugacities associated to the rotation on $S^2$, namely $\xi = e^{2 \pi i \omega}$. To be concrete, we would have:
 
\begin{align}
 \begin{split}
 Z(p_a, \Delta_a)_{\text{refined}} & = \frac{1}{N!}\sum_{m} \oint_{\mathcal{C}}\prod_{\mu=1}^{N-1} \left(d u_{\mu} \eta(q)^2\right) \mathcal{Z}_{TTref}(u, \Delta_a, \tau, p_a)\, , \label{eq:TTIref}\\
 \mathcal{Z}_{TTref}(u, \Delta_a, \tau, p_a)& =\prod_{i,j=1}^N\left[\frac{\theta_1\left(u_{ij}+ 2 \omega j; \tau\right)}{i \eta(q)}\prod_{a=1}^3 \left(\frac{i \eta(q)}{\theta_1\left(u_{ij}+ \Delta_a+ 2 \omega j; \tau\right)}\right)^{m_{ij}- p_a +1}\right]\, .
 \end{split}
\end{align} 
The refined topologically twisted index has been studied, in the strict Cardy-like limit, in \cite{Hosseini:2019lkt}, where the correction due to the refinement could be factored out in the following way:
\begin{align}
\begin{split}
Z(p_a, \Delta_a)_{\text{refined}}\big|_{\tau \rightarrow 0} &  = Z(p_a, \Delta_a) Z_{\omega}\, , \label{eq:fact}
\end{split}
\end{align}
where $Z(p_a, \Delta_a)$ is the unrefined topologically twisted index, and $Z_{\omega}$ is the correction associated to the refinement. The explicit form of $Z_{\omega}$ is irrelevant to us, while only the fact that it is independent on $u$, $p_a$ and $\Delta_a$ will be important. To the best of our knowledge, the direct application of the Bethe-Ansatz approach to the refined topologically twisted index has not been performed yet. However, we can exploit the fact that in the Cardy-like limit there is a simple connection to the unrefined index, namely \eqref{eq:fact}, and based on the intuition we have gained by studying the SCI, to argue that the combinatorial origin of $\log N$ corrections is still there at small $\tau$, therefore we do not expect it to go away as we depart from the Cardy-like limit.

As we have discussed in the AdS$_5$ black hole case, the logarithmic correction to the entropy can be seen as essentially arising from the degeneracy of dominant Bethe-Ansatz solutions to the appropriate partition function of the boundary $\mathcal{N}=4$ SYM. As in the case of the SCI, the logarithmic correction we compute for the topologically twisted index is in the grand-canonical ensemble. However, since we find that the result matches that of the microcanonical ensemble, we conjecture that there are no additional logarithmic contributions associated to the change of ensembles. Therefore, for the BPS rotating AdS$_5$ black string considered in \cite{Hosseini:2019lkt, Hosseini:2020vgl}, the logarithmic correction to $\log Z^{\text{(leading)}} (p_a, \Delta_a)$ can be obtained from the degeneracy of dominant residues contributing to the topologically twisted index of $\mathcal{N}=4$ SYM, i.e. 
\be
\Delta \log Z(p_a, \Delta_a)=\log N\, .
\ee  
This result has the same origin (in the Bethe-Ansatz treatment \cite{Benini:2018ywd}) as in the SCI, and we expect a similar robustness as the logarithmic correction to the AdS$_5$ black hole.

  Since $\textrm{log}\, Z^{\text{(leading)}} (p_a , \Delta_a)\sim N^2$ and it is homogeneous of degree one in the chemical potentials,  it is possible to apply the result of Sec.~\ref{sec:4dGCMC} to conclude that the logarithmic correction has an additional contribution from the change of ensemble which again takes the form $-d \log N$, where $d=3$ is the number of independent chemical potentials for the rotating AdS$_5$ black string.   We then conclude that 
    \begin{align}
        \Delta S_{\text{CFT}_4} & = (1-d) \log N + \mathcal{O}(1) = - 2 \log N + \mathcal{O}(1).
    \end{align}

\subsection{AdS$_5$ Black String Entropy from Near-Horizon CFT$_2$}

The near-horizon geometry of the rotating AdS$_5$ black string solution is \cite{Hosseini:2020vgl}
\begin{align}
  ds^2 & = \frac{(- \mathcal{M}\, \Pi)^{2/3}}{\Theta^2} \left[- r^2\, d\tau^2 + \frac{dr^2}{r^2} + \frac{\mathcal{W} \Theta^2}{\mathcal{M}^2} \left(dy - \frac{\mathcal{M}}{\Theta \sqrt{\mathcal{W}}}\, r d\tau \right)^2 \right] \nonumber\\
  {} & \quad + \frac{(- \mathcal{M})^{2/3}}{\Pi^{1/3}} \left[d\theta^2 + \textrm{sin}^2 \theta \left(d\varphi + \frac{\mathcal{J}}{\mathcal{M}}\, dy \right)^2 \right]\, ,\label{eq:black string NH metric}
\end{align}
where $\mathcal{M} \equiv - p_1 p_2 p_3$ is the product of magnetic charges, and ${\mathcal{J}}$ is the angular momentum, while
\begin{align}
\begin{split}
  \Theta & \equiv (p^1)^2 + (p^2)^2 + (p^3)^2 - 2 (p^1 p^2 + p^1 p^3 + p^2 p^3)\, ,\\
  \Pi & \equiv (- p^1 + p^2 + p^3) (p^1 - p^2 + p^3) (p^1 + p^2 - p^3)\, ,\\
  \mathcal{W} & \equiv \frac{-4 q_0 p^1 p^2 p^3 - \mathcal{J}^2}{\Theta}\, ,
\end{split}
\end{align}
with $q_0$ denoting the momentum added along the black string direction. Using the standard Kerr/CFT correspondence, we obtain the central charge of the near-horizon CFT$_2$
\be
  c_L = \frac{6 \mathcal{M}}{G_4 \Theta}\, . 
\ee
This central charge was found in \cite{Hosseini:2020vgl} as a Brown-Henneaux central charge \cite{Brown:1986nw}.

To compute the black string entropy using the Cardy formula, we still need the Frolov-Thorne temperature, which can be computed from the standard formalism for the Kerr/CFT correspondence \cite{Compere:2012jk}
\be
  T_L = \frac{\sqrt{\mathcal{W}}\, \Theta}{2 \pi \mathcal{M}}\, .
\ee
Therefore, the Cardy formula leads to the rotating AdS$_5$ black string (BS) entropy
\be\label{eq:AdS5 BS S Cardy}
  S_{\text{BS}} = \frac{\pi^2}{3} c_L T_L = \frac{\pi \sqrt{\mathcal{W}}}{G_4}\, ,
\ee
which is the same as the leading-order rotating AdS$_5$ black string entropy \cite{Hosseini:2019lkt, Hosseini:2020vgl}.

For the logarithmic correction to the AdS$_5$ black string entropy from the near-horizon CFT$_2$, we apply the same technique as the AdS$_5$ black hole case. As mentioned in \cite{Hosseini:2019lkt, Hosseini:2020vgl}, the rotating AdS$_5$ black string solution has one angular momentum and three electric charges. Similar to the BPS AdS$_5$ black hole case,
\be
  c\sim N^2\, ,\quad k^{11} \sim N^2\, ,\quad k^{ii} \sim N^{-2}\quad (i = 2, 3, 4)\, ,
\ee
where we take $n = 4$ in the general formula \eqref{eq:BPSdetA} due to the following reason. Three $U(1)$ electric charges have three corresponding chemical potentials $\Delta_a$ subject to a constraint, hence there are only two indepedent $U(1)$ electric charges. The angular momentum $\mathcal{J}$ appearing in the second line of \eqref{eq:black string NH metric} can be viewed as an additional $U(1)$, which can be treated in the same way as a $U(1)$ electric charge \cite{Sen:2012cj}, while in the first line of \eqref{eq:black string NH metric} there is actually another angular momentum hidden in the BTZ part of the metric. Hence, from the near-horizon region of the rotating AdS$_5$ black string there are still one angular momentum and three $U(1)$ charges (including $\mathcal{J}$), which are independent of each other. Unlike the AdS$_5$ black hole case where we can choose one of the two angular momenta, for the rotating AdS$_5$ black string the way of counting near-horizon symmetries is unique.

The reasoning of Sec.~\ref{sec:2dGCMC} can be followed in its entirety except that from the start there are only $4$ chemical potentials, one conjugate to angular momentum and three conjugate to electric charges,  obeying one constraint. This is in contrast with the AdS$_5$ black hole with $5$ chemical potentials, two conjugate to angular momenta and three conjugate to electric charges. The scalings of the Kac-Moody levels and other parameters are the same. Moreover, only one Lagrange multiplier is needed, $\lambda_{1}$, and we find that $\det H \sim N^{2}$, as in the case of the AdS$_5$ black hole with the same final result as in \eqref{eq: CFT2 result}.  Consequently,
\be
  (\textrm{det}\, \mathcal{A})_{\textrm{AdS$_5$ Black String}} \sim (N^2)^{- \frac{n+1}{2}}\, (N^2)^{\frac{n+3}{2}}\, \left(N^2 \right) = N^{4}\, ,
\ee
and the logarithmic correction to the leading-order AdS$_5$ black string entropy \eqref{eq:AdS5 BS S Cardy} is
\be
  \Delta S_{\text{CFT}_2} = - \frac{1}{2}\, \textrm{log}\, \frac{\textrm{det}\, \mathcal{A}}{(2 \pi)^{n+2}} =-2\, \textrm{log}\, N + \mathcal{O} (1)\, .
\ee

\section{Discussion}\label{sec:discussion}

In this paper we have explored logarithmic corrections to asymptotically AdS$_5$ supersymmetric extremal, rotating, electrically charged black holes and black strings. For each case we  examined the microstate counting  in the context of ${\cal N}=4$ SYM whereby it reduces to a combinatorial contribution from the space of solutions.   We also approached the logarithmic corrections to the entropy by  considering the microstate counting in the near-horizon geometry and its dual CFT$_2$, where  the logarithmic corrections arise as subleading contributions in the Cardy formula for the degeneracy of states. We found that the results from both approaches  precisely match for both AdS$_5$ black holes and rotating black strings.  It is instructive to write our boundary CFT$_4$ result as $(1-d)\log N$ to note that the logarithmic correction has two contributions, one that has a completely combinatorial origin and is rather universal, namely,  $\log N$,  while the other contribution from the change of ensemble depending on the number of independent chemical potentials of the theory,  $-d \log N$. Since we have $3$ independent chemical potentials, we obtain $-2 \log N$ as a correction to the microscopic entropy.
The calculation using the Cardy formula cannot resolve such splitting of the logarithmic correction, and only after being in microcanonical ensemble, an agreement can be found.  The reason is that the Kerr/CFT correspondence is a field-theoretic approach intrinsic to the near-horizon geometry. It is, therefore, insensitive to certain details of the UV region.

Our agreement in using the Cardy formula to its  logarithmic precision should come more as a surprise than as a foregone conclusion. There is  precedent where the Cardy formula leads to the wrong answer for logarithmic corrections \cite{Sen:2012cj}. Although the subtleties in applying the Cardy formula beyond its intrinsic regime are numerous, we expect that our positive results indicate the existence of resolutions which take into account particular properties of the spectrum  \cite{Hartman:2014oaa, Mukhametzhanov:2019pzy}.

It would be interesting to derive the logarithmic corrections directly from the macroscopic one-loop contribution in type IIB supergravity.  It is also natural to extend our near-horizon analysis to asymptotically AdS black holes in other dimensions. This route is certain to encounter obstructions in the form of zero modes, as is the case for asymptotically AdS$_4$ and AdS$_6$ black holes. Indeed, it has been shown that the one-loop supergravity contribution to the logarithmic corrections for asymptotically AdS$_4$ black holes \cite{Liu:2017vbl} is different from the one obtained in the near-horizon approach \cite{Liu:2017vll,Jeon:2017aif}. Our work indicates that given the absence of obstructions (zero modes) in odd-dimensional AdS spacetimes the counting can be performed at the near-horizon level, paving the way for a quantum entropy formula \`a la Sen \cite{Sen:2008vm}.  It will also be interesting to explore the implications of our near-horizon results within supergravity localization along the lines of \cite{Dabholkar:2014wpa,Nian:2017hac}.

\section*{Acknowledgments}

We thank  J. Hong, J. Liu and C. Uhlemann for comments and V. Godet for insightful discussions.  We also thank the referee for providing insightful comments.  This  work was supported in part by the U.S. Department of Energy under grant DE-SC0007859. M.D. is supported by the NSF Graduate Research Fellowship Program under NSF Grant Number: DGE 1256260. J.N.  thanks the Simons Center for Geometry and Physics  for warm hospitality during the initial stage of this work. A. G. L. is supported by an appointment to the JRG Program at the APCTP through the Science and Technology Promotion Fund and Lottery Fund of the Korean Government and the National Research Foundation of Korea (NRF) grant funded by the Korea government (MSIT) (No. 2021R1F1A1048531).

\appendix

\section{Special Functions} \label{app:SpecialFunctions}

Here we summarize the definitions of special functions used in the paper. The Dedekind eta function is defined as
\begin{align}
\eta(q) &  = q^{\frac{1}{24}} \prod_{k=1}^{\infty}\left( 1- q^k\right), \hspace{2mm} \text{Im}(\tau)>0\, ,
\end{align}
with $q=e^{2 \pi i \tau}$.
The Pochhammer symbol is defined as
\begin{equation}
	(z;q)_{\infty}=\prod_{k=0}^\infty(1-zq^k)\, .\label{def:pochhammer}
\end{equation}
The elliptic theta functions which are relevant to us have the following product form:
\begin{subequations}
\begin{align}
	\theta_0(u;\tau)&=\prod_{k=0}^\infty(1-e^{2\pi i(u+k\tau)})(1-e^{2\pi i(-u+(k+1)\tau)})\, ,\label{eq:theta:0}\\
	\theta_1(u;\tau)&=-ie^{\frac{\pi i\tau}{4}}(e^{\pi i\tau}-e^{-\pi i\tau})\prod_{k=1}^\infty(1-e^{2\pi ik\tau})(1-e^{2\pi i(k\tau+u)})(1-e^{2\pi i(k\tau-u)})\nonumber\\
	&=ie^{\frac{\pi i\tau}{4}}e^{-\pi iu}\theta_0(u;\tau)\prod_{k=1}^\infty(1-e^{2\pi ik\tau})\, .\label{eq:theta:1}
\end{align}\label{eq:theta}%
\end{subequations}
The elliptic gamma function and the ``tilde'' elliptic gamma function are defined as
\begin{subequations}
\begin{align}
	\Gamma(z;p,q)&=\prod_{j,k=0}^\infty\frac{1-p^{j+1}q^{k+1}z^{-1}}{1-p^jq^kz}\, ,\label{def:gamma}\\
	\widetilde\Gamma(u;\sigma,\tau)&=\prod_{j,k=0}^\infty\frac{1-e^{2\pi i[(j+1)\sigma+(k+1)\tau-u]}}{1-e^{2\pi i[j\sigma+k\tau+u]}}\, .\label{def:tilde:gamma}
\end{align}
\end{subequations}

\subsection{Asymptotic Behavior} \label{app:assypmbehavior}
For a small $|\tau|$ with fixed $0<\arg\tau<\pi$, the Pochhammer symbol can be approximated as
\begin{equation}
	\log(q;q)_\infty=-\frac{\pi i}{12}(\tau+\frac{1}{\tau})-\frac12\log(-i\tau)+\mathcal O(e^{\frac{2\pi\sin(\arg\tau)}{|\tau|}})\, .\label{pochhammer:asymp}
\end{equation}
To study asymptotic behaviors of elliptic functions, it is useful to introduce the function $\{u\}_\tau$, as
\begin{equation}
	\{u\}_\tau\equiv u- \lfloor\text{Re}( u)-\cot(\arg\tau)\text{Im} (u)\rfloor\quad(u\in\mathbb C)\, ,\label{tau-modded}
\end{equation}
which satisfies
\begin{equation}
	\{u\}_\tau=\{\tilde u\}_\tau+\check u\tau\, ,\qquad
	\{-u\}_\tau=\begin{cases}
	1-\{u\}_\tau & (\tilde u\notin\mathbb Z)\, ,\\
	-\{u\}_\tau & (\tilde u\in\mathbb Z)\, ,
	\end{cases}
\end{equation}
where we have defined $\tilde u,\check u\in\mathbb R$ as 
\begin{equation}
	u=\tilde u+\check u\tau\, .\label{u:component}
\end{equation}

The elliptic theta function $\theta_0(u;\tau)$ can be approximated for a small $|\tau|$ with fixed $0<\arg\tau<\pi$ as
\begin{equation}
\begin{split}
	\log\theta_0(u;\tau)&=\frac{\pi i}{\tau}\{u\}_\tau(1-\{u\}_\tau)+\pi i\{u\}_\tau-\frac{\pi i}{6\tau}(1+3\tau+\tau^2)\\
	&\quad+\log(1-e^{-\frac{2\pi i}{\tau}(1-\{u\}_\tau)})\left(1-e^{-\frac{2\pi i}{\tau}\{u\}_\tau}\right)+\mathcal O(e^{\frac{2\pi\sin(\arg\tau)}{|\tau|}})\, .\label{elliptic:theta:0:asymp}
\end{split}
\end{equation}
The elliptic theta function $\theta_1(u;\tau)$ is approximated for a small $|\tau|$ with fixed $0<\arg\tau<\pi$ as
\begin{equation}
\begin{split}
	\log\theta_1(u;\tau)&=\frac{\pi i}{\tau}\{u\}_\tau(1-\{u\}_\tau)-\frac{\pi i}{4\tau}(1+\tau)+\pi i\lfloor\text{Re}( u)-\cot(\arg\tau)\text{Im}( u)\rfloor+\frac12\log\tau\\
	&\quad+\log(1-e^{-\frac{2\pi i}{\tau}(1-\{u\}_\tau)})\left(1-e^{-\frac{2\pi i}{\tau}\{u\}_\tau}\right)+\mathcal O(e^{\frac{2\pi\sin(\arg\tau)}{|\tau|}})\, .\label{elliptic:theta:1:asymp}
\end{split}
\end{equation}

For a small $|\tau|$ with fixed $0<\arg\tau<\pi$, the elliptic gamma function can be approximated as
\begin{equation}
\begin{split}
	\log\widetilde\Gamma(u;\tau)&=2\pi i\,Q(\{u\}_\tau;\tau)+\mathcal O(|\tau|^{-1}e^{\frac{2\pi\sin(\arg\tau)}{|\tau|}\min(\{\tilde u\},1-\{\tilde u\})})\, ,\label{elliptic:Gamma:asymp}
\end{split}
\end{equation}
provided $\tilde u\, \slashed\to\, \mathbb Z$ (see \cite{ArabiArdehali:2019tdm} for example), and the function
 $Q(\cdot\, ; \cdot)$ is defined as:
 
 \begin{equation}
	Q(u;\tau)\equiv-\frac{B_3(u)}{6\tau^2}+\frac{B_2(u)}{2\tau}-\frac{5}{12}B_1(u)+\frac{\tau}{12}\, ,\label{eq:Q:Gamma:App}
\end{equation}
with $B_n(u)$ being the $n$-th Bernoulli polynomial.

\bibliographystyle{utphys}
\bibliography{LogCorrection}
\end{document}